\documentstyle{article}
\newcommand{\be}{\begin{equation}}
\newcommand{\hi}{\xi}
\newcommand{\ee}{\end{equation}}
\newcommand{\OX}{\overline{X}}
\newcommand{\OF}{\overline{F}}
\newcommand{\PD}{\partial}
\def\np{Nucl.\ Phys.\ }

\def\cmp{Commun.\ Math.\ Phys.\ }

\begin{document}
\begin{center}
{\Large\bf
Descendants constructed from matter field
in topological Landau-Ginzburg theories
coupled to topological gravity}\\
\vspace{0.25in}
{\large\sc A. Losev}\\
\vspace{0.25in}
{\normalsize\em
Institute for Theoretical and Experimental Physics ,Moscow\\
e-mail:Lossev@nbivax.nbi.dk;Lossev@vxitep.itep.
msk.su }

\vspace{0.25in}

{\bf
Abstract}
\end{center}

It is argued that gravitational descendants in the theory of
topological gravity coupled to
topological Landau-Ginzburg theory
  can be constructed
from  matter fields alone (without metric fields and ghosts).In this
sense topological gravity is "induced". We discuss the mechanism of this
effect(that turns out to be connected with  K.Saito's  higher
residue pairing: $K^i(\sigma_i (\Phi_1),\Phi_2)= $
$K^0(\Phi_1, \Phi_2)$ ,
and demonstrate how it works in a simplest nontrivial example:
correlator on a sphere with four marked points .
We also discuss some results on k-point correlators on a sphere.

{}From the idea of "induced" topological gravity it follows that the theory
of "pure" topological gravity (without topological matter) is equivalent
to the "trivial" Landau-Ginzburg theory (with quadratic superpotential).

%PACS Nos.
\newpage
\section{  Introduction.}
Topological theory coupled to topological gravity[1-9]
 is believed
to describe the "general"  2d matter coupled to 2d gravity.
Moreover, there is a hope that some canonical  transformation
exists that "translate" physical  theories  into  topological
one's. At the same time ,  topological  theories  coupled  to
gravity should  be  exactly  solvable.  But  still  the  full
partition  function  has  not  been  obtained  in  a  general
topological theory coupled to topological gravity. The GKM
models [10]  seems  to  give  such  a  partition  function  for
topological sigma  models  of  type  B
 [5] with  one-dimensional
target  spaces.  This  is  why  we  think  that   topological
sigma models coupled to topological gravity  deserve  further
investigation.

Here
(after a brief review in sec.2 of topological theories coupled to
topological gravity) we will argue (in sec.3) that
gravitational  descendants  in   topological   Landau-Ginzburg
theories can be constructed exclusively from matter fields
(compare this with descendants constructed from matter fields,
ghosts and/or the Liouville field in [8,11]). For example, in
Landau-Ginzburg theory with superpotential $X^n$ ,
$\sigma_{k}(X^j)= X^{j+nk}$.

{}From our construction it  follows  that  topological  gravity
itself can be constructed in terms of Landau-Ginzburg theory
with "trivial" superpotential $X^2$ [12].

In section 4 we will check arguments of sec.3 by deriving the
recursion relations[3]  for  the  four  point  correlator  on  a
sphere.
In section 5 we will discuss formula for k-point correlators and
its consequences.And in conclusion we observe that  descendants  are
connected with the higher residue pairing of K.Saito[13,14]:
\be
K^i(\sigma_i (\Phi_1),\Phi_2)=
K^0(\Phi_1, \Phi_2)
\ee

\section{  Coupling of topological matter to topological
gravity}
We will call  topological matter a 2d theory with the following
properties:

1.A theory has anticommuting symmetry $Q_m$:
\be
\{ Q_m , Q_m \}=0
\ee

2.The action of the theory can be written as:
\be
S=S_{top}+\{ Q_m , R \}
\ee
where $S_{top}$ is independent of the metric on the surface
(topological action), and
$R$ is some function
of the fields of the theory. Without the last term in (3) the functional
integral
is ill-defined, so one can say that this term regularizes topological action.
{}From   representation  (3)  one  can  easily  see   that   the
energy-momentum tensor
\be
T_{\mu \nu}=\{ Q_m, G_{\mu \nu} \}
\ee
where
\be
G_{\mu \nu}=
\frac{\delta R}{\delta g^{\mu \nu}}
\ee
The symmetry $Q_m$ acts on the space of local operators as an
external
derivative, and we call a local operator closed if it is annihilated by
$Q_m$ and exact if it is $Q_m$-derivative of some other operator.
So, the correlator of  $Q_m$-closed operators
(  according to the standard arguments [1,8]) is independent of the metric
on a surface and is  of the explicit form of regulator $R$.

Topological gravity  contains metric $g_{\mu \nu}$ and a tensor field
$\Psi_{\mu \nu}$ that is superpartner of $g_{\mu \nu}$
 in supersymmetry $Q_G$:
\be
\{ Q_G , g_{\mu \nu} \}= \Psi_{\mu \nu};
\ee
The action of topological gravity  is zero, so naively the corresponding
functional integral should be
\be
\int {\cal D} g {\cal D} \Psi ,
\ee
i.e. it is just a functional integral over the space of all metrics.
At the same time, this space of all metrics has symmetries (
reparametrizations,   conformal    symmetries    and    their
superpartners).
After fixing these symmetries  we come to the integral over the
moduli space $M$ of conformal  structures  of  metrics  on  a
surface:
\be
\int_{M} dm \int d\Psi_{0} \int {\cal D}\hi_G \exp( S_{GF}),
\ee
Here, $S_{GF}$ is an action that arises after fixing the  symmetries,
 $d\Psi_{0} $ stands for integration over supermoduli  ,$\hi_G$
stands for all  fields  except  moduli  and  supermoduli(i.e.
$\hi_G$ denotes Liouville field, ghosts,Lagrange multiplier and
their
superpartners).

The next step is coupling topological gravity to matter.  Let
us  start  with  coupling  to  matter   before   fixing   the
symmetries.
According to [8,16] it can be done as follows:
\be
S= S_{top}+\{ Q_G + Q_m , R\}= S_{m} +\int \Psi G
\ee
{}From representation (9) we see that the action describing coupled
system of metric and gravity is invariant under total supersymmetry
$Q_t =Q_G + Q_m$.

After fixing  the  symmetries  new  supersymmetry  $Q_{BRST}$
arise,and local observables $\Phi_{n}^{grav}$ in the coupled system
are $Q_t + Q_{BRST}$ cohomologies[8].
The correlator of local observables
$\Phi_{1}^{grav} , \ldots , \Phi_{n}^{grav}$
 on a Riemann surface $\Sigma$
 of genus $q$  is defined as:
\be
<\Phi_{1}^{grav} , \ldots , \Phi_{n}^{grav}>_q =
\int_{M_{q,n}} dm \mu(\Phi,m)
\ee
where the measure of integration over moduli  space  $\mu$  is
given by
\be
\mu(\Phi,m)=
\int d\Psi_{0} \int {\cal D} \hi_G {\cal D} \hi_m
 \Phi_{1}^{grav}(z_1) , \ldots , \Phi_{n}^{grav} (z_n)
 \exp( S_{GF}+ S_m +\int_{\Sigma}
 G \Psi)
\ee
Here $M_{q,n}$ is a moduli space of conformal structures on a surfaces of
genus $q$ with n punctures $z_1 \ldots z_n$, and $\hi_m$  are
fields of topological matter.

It was shown in [8] that for a topological
matter theory with nontrivial local
observables $\Phi_1  ,  \ldots  ,  \Phi_k  $   the  space  of
$Q_t+Q_{BRST}$ cohomologies
$\Phi^{grav}$ (in this theory coupled with topological
gravity)
is formed from classes  $\sigma_i (\Phi_{a}), i=0,1,2 \ldots  $ ;
$ a=1, \ldots , k$; class $\sigma_i(\Phi)$
is called the i-th descendant of the matter class
$\Phi$ and $\sigma_0(\Phi)=\Phi$.
In [8,17]
the following representatives in these classes were found:
\be
(\gamma_0)^i \Phi_a \in \sigma_i(\Phi_a)
\ee
 where  the  field  $\gamma_0$  is  constructed   only   from
ghosts,metric and their superpartners.

In the next section we will  argue  that  in  Landau-Ginzburg
topological theories coupled  to  topological  gravity  there
are  representatives  in  classes   $\sigma_i(\Phi_a)   $that
contain  only  matter  fields;  these   representatives   are
annihilated  by  both  $Q_m$  and  $Q_{BRST}$.  Using   these
representatives,  we  can  simplify  expression  (11)  for  the
measure $\mu(\Phi,m)$ by taking  the  integral  over  $\hi_G$
fields;
i.e.
\be
\mu(\Phi,m)=
\int  d\Psi_{0} \int  {\cal D}\hi_m
 \Phi_{1}^{grav}(z_1) , \ldots , \Phi_{n}^{grav} (z_n)
 \exp(  S_m +\int_{\Sigma}
 G \Psi_{0})
\ee

\section{ Local observables in the topological
Landau-Ginzburg theory coupled to topological gravity:
 appearance of "gravitational"  descendants
 constructed from matter fields. }
Topological Landau-Ginzburg(LG) theory[6,7,8,9] is a twisted $N=2$ sigma-model
of the type B (in classification of Witten[5])  on
 $C^n$ with  superfields $\hat{X_{i}}$
and $\hat{\OX_{i}}$ that are
,respectively, chiral and antichiral representations of $N=2$
 supersymmetry.
The action of such a theory is a sum of a $D$-term, $F$-term and $\OF$-term.
{}From the point of view of $Q_m$-symmetry the $D$-term and the $\OF$-term are
exact , while the $F$-term forms a topological action.
The $F$-term is constructed  from a holomorphic function on
the target space;
in LG theory this function (by definition of LG-theory) is  a
polynomial
$V(X_1, \ldots ,X_n)$.

{}From the $N=2$ SUSY one can show that among fields of the theory only
lowest components of the chiral fields are  $Q_m$-closed  and
not $Q_m$-exact;
thus, at the first glance , all polynomials of $X$
($X$ is a lowest component of the
chiral superfield $\hat{X}$) are local observables.
But the property of being a local observable does not
exclude that this observable is  zero  due  to  equations  of
motion.
Specifically,
suppose that the $D$-term describes a flat metric
on a target space and $X$ are co-ordinates
in which this metric is constant(let us call this constant $\lambda$)
, i.e. the $D$-term equals:
\be
\int_{\Sigma}\lambda \sum_{p} ( \PD X_p \PD \OX_p + fermions + \OF_p F_p )
\ee
Since the $F$-term is
\be
\int_{\Sigma}\sum_{p}\frac{\partial V}{\partial X_p}F_p + fermions
\ee
the  equations  of  motion
restrict the
axillary field $\OF_p$ to:
\be
\OF_p= (\lambda)^{-1} \frac{\PD V}{\PD X_p}
\ee
Because even off-shell $\OF_p$ is a $Q$-exact object,
$ \frac{\PD V}{\PD X_i}$ should be  zero-observable.
In a topological theory itself it would mean that for any polynomial
$P(X)$ the observable $P(X) \frac{\PD V}{\PD X_p}$ is also a zero-observable.
Really, the  correlator with insertion of
$ \frac{\PD V}{\PD X_p}$ equals to zero:
\be
<P(X)  (z)   \frac{\PD  V}{\PD  X_p}(w)  P_1(X)(w_1)   \ldots
P_k(X)(w_k)>=0
\ee
and this correlator is independent of the  positions  of  the
punctures.
That is why we can take $z=w$ and see that
$P(X) \frac{\PD V}{\PD X_p}$ is also a zero-observable.

Surprisingly, in the theory  of  LG  coupled  to  topological
gravity, the observable
$ P\frac{\PD V}{\PD X_p}$ (for a general polynomial $P$)
turns out to be a nonzero-observable.
In this theory we cannot use the argument presented above,
because the correlator (by definition) is obtained after integration
over positions of punctures($=$ over moduli
of conformal structures)  ,and  the  measure  of  integration
depends on $z$ and $w$.

Another standard
argument that naively proves the exactness of $ \frac{\PD V}{\PD X_p}$
is as follows:
if $P(X)$ is an observable constructed from the fields of matter
 (i.e. is $Q_m$-closed)   and
$\Phi$ is $Q_m$-exact  (i.e. $\Phi = \{ Q_m,\nu \}$ for some well-defined
matter field $\nu$),
then $P(X)\Phi$ is a $Q_m$-exact:
\be
P(X)\Phi=\{Q_m , P(X)\nu \}
\ee
and,thus,  a  $Q_t$  exact (because  on well-defined matter  fields
operator  $Q_t$
coincides with $Q_m$).

The point is that this argument is true only if $P(X)\nu$
is also a well-defined field.
 But this can be spoiled in the topological theory coupled to topological
gravity by the following mechanism:
 Suppose that correlator of
$P(z),\nu(w)$ in the functional integral (11)
(that gives the measure on the moduli space in topological gravity
coupled to matter) tends to infinity
as $z\rightarrow w$
(it can happen because of $G\Psi$ in the action of the theory).
Then metric field is needed to regularize the ill-defined field
$P(z)\nu(z)$ , but even this cannot be done uniformly for all metrics.

Let us see how a version of this mechanism  works in the LG-theory.
It is convenient to represent the $Q_t$-charge a sum of "holomorphic"  and
"antiholomorphic" charges(see [8,17], before twisting they corresponded
to $(dz)^{1/2}$ and  $(d\overline{z})^{1/2}$ components of the supercharge):
\be
Q_t=Q_{t,h} +Q_{t,a}
\ee
Then $\OF_p$ as the highest component of the antichiral field equals to:
\be
\OF_p = \{Q_{t,a} ,\{Q_{t,h} , X_p \} \}
\ee
Thus,
\be
X_p \frac{\PD V}{\PD X_p}=
\lambda  \{Q_{t,a} ,\{Q_{t,h} ,\OX_p  X_p \} \}
\ee
Now,  we choose such co-ordinates near the point of collision of
$X$ with $\OX$ that metric takes the diagonal form:
\be
g_{\alpha \beta}=e^{\phi} \delta_{\alpha \beta}
\ee
 The contribution from the regulator to $\OX_p  X_p$
is equal to
\be
\lambda^{-1} \phi
\ee
so we can rewrite (21) as an equality in $Q_t$-cohomology:
\be
X_i \frac{\PD V}{\PD X_i}=
  \{Q_{t,a} ,\{Q_{t,h} ,\phi \} \}=\gamma_0
\ee
( the last equality in cohomologies was obtained in [8,17])
Note that in the right side of (24) we find one of
the representations of the dilaton $\sigma_{1}(1)$ and $\lambda$
as a nonphysical parameter
  disappeared.

This effect is similar to the origin of anomalous dimension $P^2$ in
the operator $e^{iPX}$ in conformal theory of the free scalar field $X$.

Now we will restrict ourself to the simplest LG-theory:
the theory of the
one field $X$ with superpotential $V$ that is a polynomial with  the
highest power $n$(not necessarily homogeneous). Generalization
for the case of an arbitrary LG-theory seems to be obvious
(at least for polynomials $V$ with zero modality.
In LG-theory with one field it seems reasonable  that $n-1$
 observables $\Phi_a$ in topological theory are some linearly
independent polynomials with degree less than $n-1$.For example, one
can take monomials $1,X, \ldots , X^{n-2}$ as these polynomials.

Then, the first descendant of the polynomial $\Phi(X)$ should equal :
\be
\sigma_{1}(\Phi)(X)=V'(X) \int_{Z_1}^{X}\Phi(Y)dY
\ee
Note that changing the integration constant $Z_1$ results in adding to
$\sigma_{1}(\Phi)(X)$ a term $\Delta Z_1 V'(X)$ that is a $Q_t$-exact
object also in topological gravity coupled to LG  theory.

It is easy  to obtain higher descendants. For example,
in $Q_t$ cohomology:
\begin{eqnarray}
V'(X) \int_{Z_1}^{X}YV'(Y)dY &=& \{Q_{t,a} ,\{Q_{t,h} ,\phi \} \} X V'(X)=
 \nonumber \\
=(\{Q_{t,a} ,\{Q_{t,h} ,\phi \} \})^2 &=& \sigma_2 (1)
\end{eqnarray}
Reasoning in  this  way,  we  get  the  general  formula  for
descendants:
\be
\sigma_{m}(\Phi)(X)=
V'(X)\int_{Z_1}^{X}V'(Y_1)dY_1 \int_{Z_2}^{Y_1} \ldots
V'(Y_{m-1}) \int_{Z_m}^{Y_{m-1}}
\Phi(Y_m)dY_m
\ee
and in $Q_t$-cohomologies this expression does not depend on  $Z_i$.

The idea that matter can be responsible for gravitational dressing
can be also found in Krichever paper[15].
\section{ The four point  formula  in  LG  theory and  recursion
relations}

In this  section we will study correlators in LG theory with superpotential
$V$ coupled
to topological
gravity in genus 0,(i.e. the worldsheet is a sphere with punctures)
$<P_1(X) \ldots P_r(X) >_{V}$,
 that are integrals over the moduli space $M_{0,r}$ of conformal
structures on a sphere with $r$ punctures. Since the
complex dimension of $M_{0,r}$ is equal to $r-3$, the first
nonzero correlator appears  at  $r=3$.  This  correlator  was
shown [8,9]
to be equal to correlator in LG theory itself:
\be
<P_1 P_2 P_3>_{V}=<P_1(z_1) P_2(z_2) P_3(z_3) >_{V}^{LG}=
\int \frac{P_1(X) P_2(X) P_3(X) dX}{V'(X)}
\ee
and the integral is taken around  infinity in the complex plane .

When $r=4$ the space $M_{0,4}$ is a sphere. This can be easily seen
if we consider the space of complex structures on a sphere with 4
marked points. The projective ($SL(2,C)$) invariance allows us to
fix the co-ordinates  of the first three points and co-ordinate $z_4$
of
the fourth point can be treated as a coordinate on $M_{0,4}$. The
only difference comes when the "running" point $z_4$ "collides"
with one of
the "fixed" points(for example,
$z_4=z_1$). In the naive picture when such a collision takes
place ,fixed and running points just coincide, but in the space of
conformal structures such a collision corresponds to degeneration of a
sphere into two spheres connected by an infinitely long tube.

It was shown in [8] that at a general point in the moduli space
($z_4 \neq z_i, i=1,2,3$) the measure $\mu(\Phi,z_4)$ equals
the correlator in LG-theory of three local topological observables
and the $F$-term of the chiral superfield $P_4(X)$ :
\be
\mu(\Phi,z_4)=
<P_1(z_1) P_2(z_2) P_3(z_3) P_{4}^{(2)}(z_4)>_{V}^{LG}
\ee
where
\be
P_{4}^{(2)}(z)=
 \int d^2 \theta_{+}P_4(\hat{X})(z,\theta_{+})
\ee
Integrating $\mu$ over $M_{0,4}$(i.e. over $z_4$)
 we see that the four point correlator in topological gravity
coupled to LG-theory looks very similar to the variation of the three
point correlator (28) in LG-theory itself. Naively, if degree of
$P_4$ is less than $n$ (degree of $V$) one could think that
\be
<P_1 P_2 P_3 P_{4}>_{V}=\frac{d}{dt}<P_1 P_2 P_3>_{V+t P_{4}},\; \;
 at \; \; t=0
\ee
but this is not so because we suppose that expression (29)  for measure
$\mu$ is not correct when $z_4=z_i,\; i=1,2,3$.
It is possible to correct the situation. We do not know what really happens
when $z_4$ is near $z_i$, but we know that in that region of the moduli
space there is a very long tube between these two points and other points,
and only states corresponding to topological observables can propagate through
it.So, we can say that the difference between the  naive consideration and
what really happens is equivalent to the insertion of some local topological
observable,that is bilinear in fields $P_1$ and $P_4$. We will call such
an observable "contact term" and denote it as $C_V(P_4,P_1)$.

 Thus, the corrected
expression  of four point correlator takes the following form:
\begin{eqnarray}
&&<P_1 P_2 P_3 P_4>=
\int \frac{dX}{V'(X)} (-\frac{P_1(X) P_2(X) P_3(X) P_{4}'(X)}{V'(X)}
\nonumber \\
&&+C_V(P_4,P_1)P_2 P_3+C_V(P_4,P_2)P_1 P_3
+C_V(P_4,P_3)P_1 P_2)
\end{eqnarray}

Note that without "contact terms"  the  expression for
four  point  correlator (32)
fails to be symmetric under permutations of observables,  but
by the definition the correlator should be symmetric.

The explicit expression for the contact terms is given by the
following construction.
For a polynomial $S$ of arbitrary degree we will define a  set
of polynomials $\{ S^{(i)} \}$ $i=0,1,\ldots $ ;
$deg S^{(i)} < deg V'=n-1 $ :
\begin{eqnarray}
S&=&S^{(0)}+\frac{dV}{dX} Q^{(0)} \nonumber \\
 &\cdots&                         \nonumber \\
\frac{dQ^{(i)}}{dX}&=&S^{(i+1)}+\frac{dV}{dX} Q^{(i+1)}
\end{eqnarray}
We will take
\be
C_V(P_1,P_2)=
=\frac{d}{dX}(\frac{P_1 P_2}{V'})_{+}
=(P_1 \cdot P_2)^{(1)}+V'\cdot(something)
\ee
Here and below subscript $+$ stands for the positive part  in expansion
in $X$.
Then,the  symmetry under permutations  of  the  four  point
function  trivially follows from the identity: for arbitrary
polynomials $S_1, S_2$
\be
K^{(1)}(S_1,S_2)
\equiv \int \frac{S_{1}^{(1)} S_2 -
S_{2}^{(1)} S_1}{V'}=
\frac{1}{2} \int \frac{S_1 ' S_2 -S_2 ' S_1}{(V')^2}
\ee
Note that $K^{(1)}$ in (35) is just the  first  higher  residue
pairing of K.Saito[13,14].

We understand four point formula (32) with contact terms given by (34) was
implicitly assumed in [8].

It is interesting that while our derivation of the four point
formula is valid only for
observables $P_i$ with $deg P_i < n-1 $, the expression (32)
is symmetric under permutations and looks "natural" for all
observables. Thus, we claim that four point formula (32) is
true for all observables (it would be interesting  to  derive
it directly from the functional integral in LG theory).

The symmetry of the four-point function does not determine term
$something$ in the second representation in (34),
 but in the next section considering k-point correlators
 we will confirm this term.

Let us check  ,  using  (32),  the  recursion
relations for four point correlators. The recursion relations
derived in [3] for an arbitrary topological theory coupled to
topological gravity state that:
\be
<\sigma_1(\Phi_1) \Phi_2 \Phi_3 \Phi_4>=
\sum_{a}<\Phi_1 \Phi_2 \Phi_a>
<\Phi_{a^{*}} \Phi_3 \Phi_4>
\ee
Here local observable in topological theory $\Phi_{a^{*}}$ is conjugated to
observable $\Phi_a$ in metric $\eta$ given by the two-point correlator in
topological theory itself. In the case of LG topological theory
\be
<\Phi_{a^{*}} \Phi_{b}>_{V}^{LG}=
\int \frac{\Phi_{a^{*}}(X) \Phi_{b}(X) dX}{V'}=\delta_{a^{*}b}
\ee

Thus, it is easy to show that in LG theory coupled to gravity
\be
<\sigma_1(\Phi_1) \Phi_2 \Phi_3 \Phi_4>_{V}=
\int \frac{\Phi_1(X) \Phi_2(X) \Phi_3(X) \Phi_4(X) dX}{V'}
\ee
At the same time, substituting $\sigma_1(\Phi_1)$ in the form
of (25) into four  point  formula  (32)  ,  we  obtain  the  same
expression! This checks the consistency of our results.

\section{The $k$-point formula and its consequences
(results of this section were obtained in collaboration with
I.Polyubin)}
 \subsection{Recurrent definition of the $k$-point correlators}
Leading by the idea that the integration over the position of the marked
point (with insertion of a primary field in it)
 results in infinitesimal change of the superpotential plus contact terms,
 we can give the following recurrent definition of the $k$-point formula.

 1.3-point formula is given by the integral (28).

 2.If all fields in the correlators are gravitational descendants of
 primary fields, correlator is equal to zero. Really, if it is the case
 then the degree of form to be integrated over moduli space is at least
 $2k$ (due to at least $2k$ $\gamma_0$) while the
 real dimension of the corresponding
 moduli space(in genus zero) is $2k-6$.

 3.If one of the insertions(let it be $P_k$ for definiteness) is a primary
 field, then
 \begin{eqnarray}
 &&<P_{1}\ldots P_{k}>_{V}=
 \frac{d}{dt}<P_{1}\ldots P_{k-1}>_{V
 +tP_{k}}|_{t=0}+  \nonumber \\
 &&+ \sum_{i=1}^{k-1}<P_{1}\ldots C_V (P_{i},P_{k})
 \ldots P_{k-1}>_{V}
 \end{eqnarray}

One can  check, that this definition is consistent in the
following sense.
 Suppose that
there are,for example, two primary fields $\Phi_i$
and $\Phi_j$ in $k$-point correlator. Reducing to $(k-1)$-point
correlator using $\Phi_i$ and then reducing once again
 to $(k-2)$-point correlator
using $\Phi_j$(and the contact term between them) one gets the same result as
if one starts with $\Phi_j$ and then proceeds with $\Phi_i$.

\subsection{Puncture and dilaton equations}

Operator that is equal to $1$ is called puncture operator(this name
tells us that this $1$ should be placed at a marked point,i.e. puncture).
In order not to confuse it with a number we will denote it as $1_{P}$.
Reduction (39)
 with the help of the puncture does not change the derivative of
the superpotential $V$  and constant shift in $V$ cannot be seen neither in
subsequent contact terms , nor in the final 3-point function, so the only
effect of puncture are contact terms that are:
\be
C_V(1_{P},\sigma_{n}(\phi_{i}))=\sigma_{n-1}(\phi_{i})
\ee
Thus, for the correlator of puncture with some other fields we have the so
called puncture or $L_{-1}$ equation:
\be
 <1_{P}\sigma_{n_{1}}(\Phi_1) \ldots\sigma_{n_{k}}(\Phi_k)>=
 \sum_{i=1}^{k}
 <\sigma_{n_{1}}(\Phi_1)\ldots\sigma_{n_{i}-1}(\Phi_i)\ldots
 \sigma_{n_{k}}(\Phi_k)>
\ee
It is also possible(using induction in number of fields in the correlator)
to prove the dilaton equation:
\be
<\sigma_{1}(1) P_{1} \ldots P_{k}>=(k-2)<P_{1}\ldots P_{k}>
\ee
\subsection{Correlator $<
\sigma_{n_1}(\Phi_{i_1}) \ldots \
\sigma_{n_k}(\Phi_{i_k}) >$ for $\sum n_{\alpha}=k-3$ }

This case is distinguished because here the total
dimension of all first Chern classes $\gamma_0$ is equal to dimension
of the moduli space. Inductively, using rules described above, one can show
that:
\be
<\sigma_{n_1}(\Phi_{i_1})\ldots\sigma_{n_k}(\Phi_{i_k})>=
\frac{(n_{1}+\ldots+n_{k})!}{n_{1}!\ldots n_{k}!}\int dX
\frac{\Phi_{i_1}\ldots\Phi_{i_k}}{V'}
\ee

The
 first factor in (43) corresponds to the contribution coming purely
from the topological theory itself, while the second is nothing but
correlator in LG topological theory itself.
 Such representation is expected on general grounds,
see, for example,[3,4,8].

In particular, from (43) it follows that correlators in $V=X^2$
theory coincide with the wellknown correlators in topological gravity
itself in genus zero, as we expected from the very beginning(sec.3).

\subsection{Introduction of "times"
on "small phase space" and so called flat coordinates}
The procedure of reduction from $k$-point correlator to $(k-1)$-point
correlator can be interpreted as an infinitesimal evolution
of $(k-1)$-point correlator caused by $\Phi_k$. In this evolution
not only superpotential is shifted , but also operator change due to
contact terms, namely:
\be
\delta_{\Phi_k}(\Phi_{i})=C_{V}(\Phi_{i},\Phi_{k})
\ee
Consistency of the reductions mentioned above imply that these evolutions
commute and it is reasonable to consider the full hierarchy of evolutions,
and to integrate it. So for a while we will deal with the following object
(global version of (39)):
\be
<\Phi_1 (t) \ldots \Phi_k (t) >_{V(t)}=
<\Phi_1 \ldots \Phi_k exp \sum t_{i}\Phi^{i}>_V
\ee
In [4,8] parameters $t$ were called "times" on the "small phase space",
exponential in (45) should be considered as a formal seria in insertions
(like in [3]).
Taking the derivative with respect to  $t$ we have:
\be
\frac{\partial}{\partial t_i} V(t) = \Phi_i(t)
\ee
\be
\frac{ \partial \Phi_k (t)}{\partial t_{i}}=
C_{V(t)}(\Phi_{i} (t),\Phi_{k}(t))
\ee

Moreover, using manifest representation of the contact term , it is possible
to show that if
\be
\Phi_{r}=\frac{d}{dX}(V^{\frac{r+1}{n}})_{+}
\ee
then
\be
\Phi_{r}(t)=\frac{d}{dX}((V(t))^{\frac{r+1}{n}})_{+}
\ee
Now it easy to explain the notion of so called flat co-ordinates
 on the space of polynomials of order n. Correlator
  $<\Phi_i(t) \Phi_j(t) 1_P>_V$  can be
  interpreted as a metric on this space. Since
  \be
  <\Phi_i(t) \Phi_j(t) \Phi(t) 1_P>_V(t)=0
  \ee
  for all $\Phi$ , as it follows from (39), the metric in
  co-ordinates $t$:
    \be
   \eta_{ij}= < \frac{\partial V(t)}{\partial t_i}
   \frac{\partial V(t)}{\partial t_j} 1_P>_V(t)
   \ee
    is $t$ independent, i.e.
   flat, and it is the set of co-ordinates $t$
    in which this metric is not only flat
   but also constant.

\section{Concluding remark: K.Saito higher residue pairing
and descendants}
One amusing thing in conclusion. From (27) and  (33)  it  follows
that
\be
(\sigma_i(\Phi))^{(i)}=\Phi
\ee
Since in the case of one variable  the  $i$-th  higher  residue
pairing of K.Saito is equal to[13,14]:
\be
K^{(i)}(P,\Phi)=      K^{(0)}(P^{(i)},\Phi)      =       \int
\frac{P^{(i)}\Phi dX}{V'}
\ee
we proved the connection (1) between  descendants  and  higher
residue  pairing.  One  can  show  that  this  connection  is
true also for LG theory in  multi-dimensional target space.

\vspace{1cm}

The author is grateful to M.Shifman and V.Zakharov (whose papers [18]
of 1983! implied many of the ideas of what is now called topological
theories) for a lot of valuable discussions on this subject.

The author is grateful to A.Gerasimov, A.Morozov, A.Rosly  and,
especially, to A.Marshakov, A.Mironov, M.Olshanetsky for critical
comments.

\vspace*{\fill}
\end{document}